\documentclass[10pt]{article}

\usepackage{amsmath}
\usepackage{amssymb}
\usepackage{graphicx}
\usepackage{psfrag}
\usepackage{color}
\usepackage{sistyle}
\usepackage{setspace}

\DeclareMathOperator{\SHO}{\mathrm{SHO}}
\DeclareMathOperator{\Sader}{\mathrm{Sader}}

\newcommand{\drawat}[3]{\makebox[0pt][l]{\raisebox{#2}{\hspace*{#1}#3}}}

\renewcommand{\Re}{\mathrm{Re}}
\renewcommand{\Im}{\mathrm{Im}}

\definecolor{linkcolor}{rgb}{0,0,1}

\usepackage[colorlinks=true, pdfstartview=FitV, linkcolor= linkcolor, citecolor= linkcolor, urlcolor= linkcolor, hyperindex=false]{hyperref}

\begin{document}

\title{\bf Frequency dependence of viscous and viscoelastic dissipation in coated micro-cantilevers from noise measurement}

\author{P. Paolino, L. Bellon\\
Universit\'e de Lyon \\
 Laboratoire de Physique, \'Ecole Normale Sup\'erieure de Lyon \\
C.N.R.S. UMR5672 \\
46, All\'ee d'Italie, 69364 Lyon Cedex 07, France\\}

\maketitle

\begin{abstract}
We measure the mechanical thermal noise of soft silicon atomic force microscopy cantilevers. Using an interferometric setup, we obtain a resolution down to $\SI{e-14}{m/\sqrt{Hz}}$ on a wide spectral range ($\SI{3}{Hz}$ to $\SI{e5}{Hz}$). The low frequency behavior depends dramatically on the presence of a reflective coating: almost flat spectrums for uncoated cantilevers versus $1/f$ like trend for coated ones. The addition of a viscoelastic term in models of the mechanical system can account for this observation. Use of Kramers-Kronig relations validate this approach with a complete determination of the response of the cantilever: a power law with a small coefficient is found for the frequency dependence of viscoelasticity due to the coating, whereas the viscous damping due to the surrounding atmosphere is accurately described by the Sader model. 
\end{abstract}

\vspace{10mm}

\noindent {\bf Keywords}

\noindent Thermal noise, viscoelasticity, cantilever coating, fluctuation dissipation theorem, Sader model

\vspace{5mm}

\noindent {\bf Notes}

\noindent This paper has been published as

\emph{Nanotechnology} {\bf 20}, 405705 (2009).

\href{http://dx.doi.org/10.1088/0957-4484/20/40/405705}{doi: 10.1088/0957-4484/20/40/405705}

\vfill

Corresponding author : Ludovic.Bellon@ens-lyon.fr

\newpage

\section{Introduction}

The interest to study micro-cantilevers, at first related to its use in atomic force microscopy (AFM), is even increased as it represent a fundamental part of microelectromechanical systems (MEMS). Functionality of MEMS or AFM sensors is based on mechanical movements and deformation of the cantilever. Their thermal noise represent one of the most important noise sources and finally determines the ultimate deflection sensitivity of the sensor \cite{Saulson-1990,Butt-1995,Yasumura-2000,Djuric-2000,Lifshitz-2000,Lavrik-2004,Lochon-2005}. This mechanical-thermal noise can also be important in the study of macroscopic systems, and has been shown for instance to be a relevant term in the sensitivity limitations of interferometric detectors for gravitational waves \cite{Gonzalez-1995,Kajima-1999,Numata-2003,Rowan-2005}. Several observations have shown that surface effects can play a significant part in understanding the origin of these thermal fluctuations \cite{Lochon-2005,Numata-2003,Rowan-2005,Crooks-2002,Datskos-2004,Sandberg-2005}.

As shown by the fluctuation dissipation theorem (FTD) \cite{Callen-1952}, these thermal induced mechanical fluctuations are linked to the losses of energy occurring during deformations of the system. Many models have been proposed to account for the numerous physical sources of dissipation: viscous damping in the surrounding fluid \cite{Sader-1998}, clamping losses \cite{Yasumura-2000}, thermoelastic dissipation \cite{Zener-1937,Zener-1938a,Zener-1938b}, etc.
In his pioneering paper Saulson \cite{Saulson-1990} proposed a model of mechanical-thermal noise for a simple harmonic oscillator with viscoelastic damping. In particular he showed that, considering only structural damping as the dissipation mechanism, the power spectrum density (PSD) of fluctuations presents a $1/f$ trend at low frequencies. This is the \emph{viscoelastic model}. In general, a key difference between all these models is the frequency dependence of the noise or dissipation. It is however a great challenge to measure thermal noise or small damping on a wide range of frequency, and very few experiments \cite{Gonzalez-1995,Kajima-1999,Numata-2003,Yamamoto-2001} have succeed so far in directly measuring fluctuations out of resonances, notably at low frequency.

We propose here a direct measurement of a mechanical-thermal noise of a microcantilever, realized with an interferometric technique. A principal advantage of our setup is that offers, thanks to its sensitivity, the possibility to resolve not only the resonances of the PSD as in the standard AFM optical lever technique, but the whole spectrum from  $\SI{3}{Hz}$ to $\SI{4e4}{Hz}$. In this paper, we will first present a short introduction to a few useful models for the cantilever noise: simple harmonic oscillator with viscous damping, viscoelastic model, Sader model. We will then present measurements on a raw silicon cantilever and a golden coated one, with clear $1/f$ like noise for the latter. Vacuum experiments to reduce the viscous effects due to the surrounding atmosphere will demonstrate that this behavior is an intrinsic property of the cantilever. With the use of FDT and Kramers-Kronig relations, we will eventually rebuild the full mechanical response function from the measured PSD. We will then propose a phenomenological model to closely mach the observations, adding a frequency dependent viscoelasticity for inner dissipation to the Sader's approach for surrounding atmosphere.

\section{Thermal noise of a damped harmonic oscillator}

When a mechanical system is in equilibrium with a thermal bath at temperature $T$, there is a continuous exchange between the mechanical energy accumulated in the system and the thermal energy of the environment. The thermal fluctuations of an observable $d$ are described by the Fluctuation-Dissipation Theorem \cite{Callen-1952}, which relates the power spectrum density (PSD\footnote{In this paper we will use one sided power spectrum density functions of frequency $f$, such that $<d^{2}>=\int_{0}^{\infty}S_{d}(f)df$, for easier comparison to experiments, whereas response functions $G$ will be given as a function of the pulsation $\omega$.}) $S_d(f)$ to the response function $G$ of the system:
\begin{equation}\label{eq:FDT}
S_d(f) =\frac{\langle d^2\rangle}{\Delta f}=-\frac{4k_BT}{\omega}\Im\left[\frac{1}{G(\omega)}\right]
\end{equation}
where $k_B$, $\Delta f$ and $\omega=2\pi f$ are the Boltzmann constant, the spectral bandwidth and the pulsation corresponding to frequency $f$. $\Im$ stand for the imaginary part of its argument. $G(\omega)$ is defined as
\begin{equation}\label{G}
G(\omega)=\frac{F(\omega)}{d(\omega)}
\end{equation}
where  $F$ is the variable coupled to $d$ in the Hamiltonian of the system.

Let us consider the case of a simple harmonic oscillator ($\SHO$) with viscous damping as a simple model of
a mechanical dissipative system, where $d$ and $F$ are the displacement and the force applied to the system. It responds to the equation of motion:
\begin{equation}
m\ddot{d}=- \kappa d -\gamma \dot{d}+ F
\end{equation}
where $m$ is the mass, $\kappa$ the spring constant and $\gamma$ the damping coefficient. In Fourier space this equation can be rewritten as
\begin{equation}\label{eq:G-SHO}
G^{\SHO}(\omega)=\kappa\left[1-\frac{\omega^2}{\omega_0^2}+i\frac{\omega}{Q\omega_0}\right]
\end{equation}
where we introduced the resonant pulsation $\omega_0=\sqrt{\kappa/m}$ and the quality factor $Q=m \omega_0/\gamma$. In this case, from equations \ref{eq:FDT} and \ref{eq:G-SHO}, the PSD of thermal fluctuations is given by
\begin{equation}\label{eq:S-SHO}
S_d^{\SHO}(f) =\frac{4k_BT}{\kappa\omega_0}\frac{1/Q}{\left(1-u^2\right)^2+(u/Q)^2}
\end{equation}
where $u=\omega/\omega_0$ is the reduced frequency.

Up to now, we have considered only viscous damping (proportional to velocity). In a more general case we can consider another dissipation source: the spring itself may be viscoelastic. This can be modeled by a complex spring constant \cite{Saulson-1990} $\kappa^*=\kappa(1+i  \phi)$ in the Fourier space. The imaginary part of complex spring constant $\kappa\phi$ takes into account the dissipation because it includes a component of the restoring force which is out of phase with the displacement. As we are interested in resonant systems, we will generally assume that $\phi \ll 1$. If we only consider viscoelastic damping, eq. \ref{eq:G-SHO} and \ref{eq:S-SHO} turn into:
\begin{align}\label{eq:G-k*}
G^{\kappa ^{*}}(\omega)&=\kappa\left[1-\frac{\omega^2}{\omega_0^2}+i\phi\right] \\
\label{eq:S-k*}
S_d^{\kappa ^{*}}(f)&=\frac{4k_BT}{\kappa\omega_0}\frac{\phi/u}{(1-u^2)^2+\phi^2}
\end{align}
In this model the resonance can be characterized by an effective quality factor $Q^{*}=1/\phi$. If we consider both the viscoelastic and the viscous damping, we have:
\begin{align}\label{eq:G-SHO*}
G^{\SHO^{*}}(\omega)&=\kappa\left[1-\frac{\omega^2}{\omega_0^2}+i\left( \frac{\omega}{Q\omega_0}+\phi \right)\right] \\
\label{eq:S-SHO*}
S_d^{\SHO^{*}}(f)&=\frac{4k_BT}{\kappa\omega_0}\frac{1/Q+\phi/u}{(1-u^2)^2+(u/Q+\phi)^2}
\end{align}
The effective quality factor $Q_{\textit{eff}}$ of the resonance is here a combination of both dissipation processes:
\begin{equation}\label{eq:Qeff}
\frac{1}{Q_{\textit{eff}}}=\frac{1}{Q}+\frac{1}{Q^{*}}
\end{equation}
If we reduce viscous damping to zero ($Q\rightarrow\infty$), $Q^{*}$ represent the upper bound of achievable effective quality factor.

The models with and without viscoelasticity present significant differences at low frequencies: the viscous damping model (labelled $\SHO$) produces a constant spectrum, while the viscoelastic models (labelled $\kappa^{*}$ and $\SHO^{*}$) give rise to a $1/f$ trend (for a frequency independent $\kappa^*$, a common observation for many material \cite{Kimball-1927,Walther-1935,Saulson-1994,Speake-1999} usually referred to as \emph{strucutral damping} is the literature). They can be useful to study the thermal noise driven fluctuations of a microcantilever around and below its first resonance. Indeed, the expansion theorem states that the response of a continuous system to an applied force is equal to the superposition of the responses of each of the normal mode of the system \cite{Saulson-1990}, hence it is usual to modelize the first mode as a simple harmonic oscillator. This classic assumption has been validated by Sader \cite{Sader-1998} around resonances with high quality factors, which is common for cantilevers in vacuum or air. Nevertheless, the Sader model \cite{Sader-1998} is much richer to describe the off resonance behavior and will be useful to understand our measurements, hence we will quickly recall its main lines here.

When a cantilever is moving inside a fluid, it is subject to a force corresponding to the hydrodynamic load  $F_{\mathrm{hydro}}(\omega)$ due to the motion of the fluid  around the beam. Following Sader's approach \cite{Sader-1998}, this hydrodynamic load can be approximated by
\begin{equation} \label{eq:F_hydro}
F_{\mathrm{hydro}}(\omega) = \frac{\pi}{4} \rho \omega^{2} b^{2} \Gamma(\omega) d(\omega)
\end{equation}
where $\rho$ is the density of the fluid and $\Gamma(\omega)$ the hydrodynamic function corresponding to a thin rectangular cantilever beam of width $b$ and length $L$. An explicit formula of $\Gamma$ for a infinitely thin cantilever much longer than wide is given in ref.~\cite{Sader-1998}. Considering only the first resonance and measuring the deflection $d$ at the free extremity of the cantilever, it can be shown that the response function $G^{\Sader}(\omega)$ is \cite{Bellon-2008}
\begin{equation}\label{eq:G-Sader}
G^{\Sader}(\omega)=1.03 \kappa_{c} -\frac{1}{4}m_{c}\left(1+\tau(\omega)\right) \omega^{2}
\end{equation}
with $\kappa_{c}$ and $m_{c}$ the static spring constant and mass of the cantilever, and $\tau(\omega)=\pi \rho b^{2} L \Gamma(\omega) /4 m_{c}$. The real part $\tau_{r}$ of $\tau$ corresponds to the added mass due the fluid moving along with the cantilever during its motion (normalized to the cantilever mass $m_{c}$), and the imaginary part $\tau_{i}$ to the viscous drag. The correspondence to the $\SHO$ model is immediate : in vacuum ($Q\rightarrow\infty$ and $\tau(\omega)=0$), we identify the effective mass of the oscillator to a quarter of the cantilever mass ($m=m_{c}/4$) and the spring constant to $\kappa=1.03 \kappa_{c}$. 
Reporting the above expression of $G^{\Sader}(\omega)$  into eq.~\ref{eq:FDT} directly gives the expected spectrum in this model:
\begin{equation}\label{eq:Sd-Sader}
S_d^{\Sader}(f) =\frac{4k_BT}{\kappa\omega_0} \frac{\tau_{i}(\omega) u}{(1-(1+\tau_{r}(\omega))u^{2})^{2} +\tau_{i}^{2}(\omega)u^{4}}
\end{equation}
The low frequency behavior of this model is a little different from the SHO model: instead of a constant spectrum at low frequency, the noise is expected to slowly vanish as $f \rightarrow 0$ \cite{Bellon-2008}.

Just as for the simple viscoelastic model ($\SHO^{*}$), one can include an internal dissipation in the Sader model by considering a complex spring constant $\kappa^{*}$ (corresponding to a complex young modulus). This extended Sader model will be labelled as $\Sader^{*}$:
\begin{align}\label{eq:G-Sader*}
G^{\Sader^{*}}(\omega)&=\kappa(1+i\phi) -m(1+\tau_{r}(\omega)+i \tau_{i}(\omega)) \omega^{2} \\
\label{eq:S-Sader*}
S_{d}^{\Sader^{*}}(f)&=\frac{4k_BT}{\kappa\omega_0} \frac{\tau_{i}(\omega) u + \phi/u}{(1-(1+\tau_{r}(\omega))u^{2})^{2} +(\tau_{i}(\omega)u^{2}+\phi)^{2}}
\end{align}
This last model will be equivalent to the simple viscoelastic oscillator ($\kappa^{*}$) at low frequencies and to the Sader model at higher frequencies. 

\begin{figure}
\begin{center}
\psfrag{Re(G)/k}[Bc][Bc]{$\Re(G)/\kappa$}
\psfrag{Im(G)/k}[Bc][Bc]{$\Im(G)/\kappa$}
\psfrag{Normalized PSD}[Bc][Bc]{Normalized PSD $S_{d}(u)$}
\psfrag{Normalized frequency u}[Bc][Bc]{Normalized frequency $u$}
\psfrag{(a)}[Bc][Bc]{(a)}
\psfrag{(b)}[Bc][Bc]{(b)}
\psfrag{(c)}[Bc][Bc]{(c)}
\psfrag{SHO}[Bl][Bl]{\small $\SHO$}
\psfrag{SHO*}[Bl][Bl]{\small $\SHO^{*}$}
\psfrag{Sader}[Bl][Bl]{\small $\Sader$}
\psfrag{Sader*}[Bl][Bl]{\small $\Sader^{*}$}
\psfrag{k*}[Bl][Bl]{\small $\kappa^{*}$}
\includegraphics{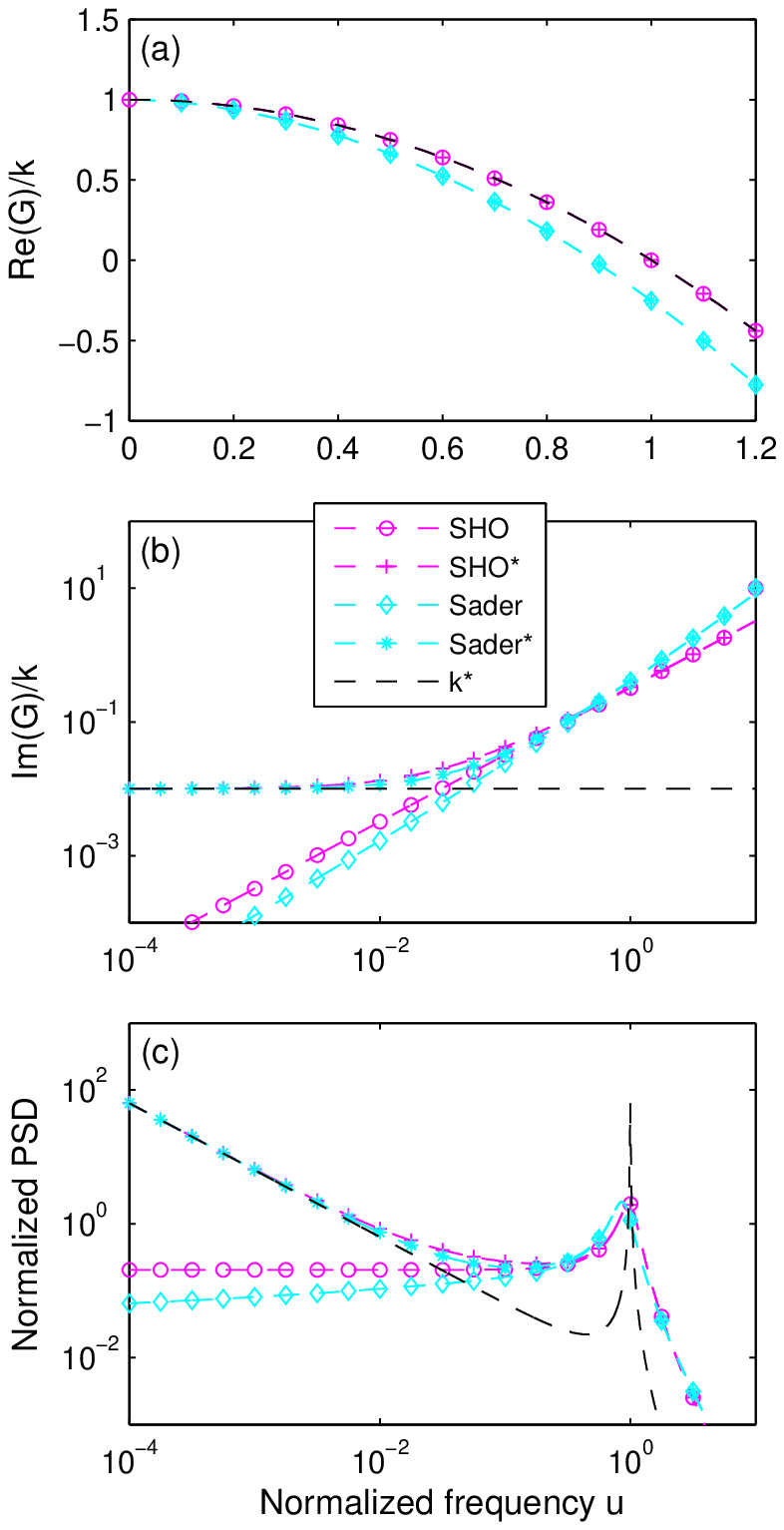}
\end{center}
\caption{Mechanical response function and thermal noise spectrum of a cantilever for various models. (a) Real part of the response function normalized by the spring constant. (b) Imaginary part of the response function normalized by the spring constant (log scale on both axes). (c) Power Spectrum Density (PSD) of thermal noise induced fluctuations, normalized by $k_{b}T/\kappa$ (log scale on both axes). Viscoelastic models prevent dissipation to vanish at low frequencies, thus raising the thermal noise in this limit. Sader models take into account the additional inertia due to the surrounding fluid, shifting the resonances to lower frequencies.}
\label{Fig:Models}
\end{figure}

We plot in Fig.~\ref{Fig:Models} the typical behavior we can expect from the 5 different models. The presence of a viscoelastic behavior is evidenced at low frequency by a none vanishing imaginary part of the response function $G$, which induces a divergence of the power spectrum of fluctuations when $f$ goes to $0$. The resonance is also influenced by the chosen model: viscoelasticity alone will produce sharper resonance, and the additional inertia due the fluid taken into account with Sader model shifts it to lower frequencies.

\section{Experiments and results}

\begin{figure}
\begin{center}
\psfrag{PSD}[Bc][Bc]{PSD $S_{d}(f)/\SI{}{(m^2/Hz)}$}
\psfrag{f}[Bc][Bc]{Frequency $f/\SI{}{Hz}$}
\psfrag{(a)}[Bl][Bl]{\small (a) BS-Cont cantilever}
\psfrag{(c)}[tl][tl]{}
\psfrag{(b)}[Bl][Bl]{\small (b) BS-Cont-GB cantilever}
\includegraphics{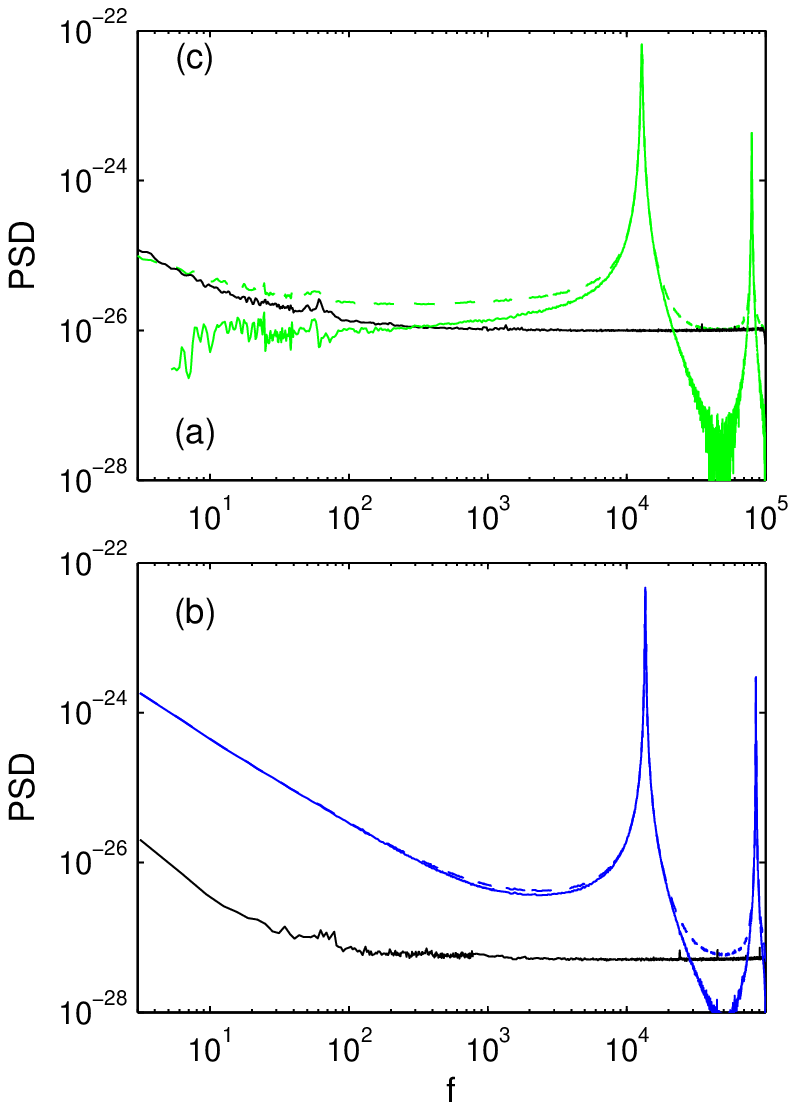}%
\drawat{-66mm}{85mm}{
\psfrag{d}[Bl][Bl]{\small $d$}%
\psfrag{Er}[Bl][Bl]{\small $E_{\mathrm{ref}}$}%
\psfrag{Es}[Bl][Bl]{\small $E_{\mathrm{sens}}$}%
\includegraphics[width=3cm]{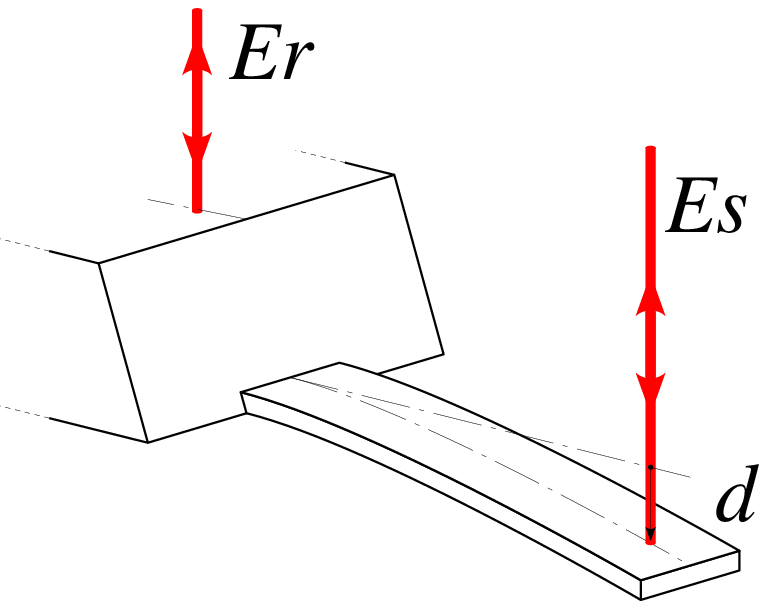}}%
\end{center}
\caption{Power Spectrum Density (PSD) of thermal noise induced fluctuations for a raw silicon cantilever (a) and a golden coated one (b). The background noise of the system (bottom black line) is measured on a rigid mirror with light intensities on the photodiodes tuned exactly as during the measurement on the cantilever. We subtract this noise from the raw measurement (dashed line) to estimate the actual thermo-mechanical noise of the cantilever (plain line). In this frequency window, the first 2 flexural resonances are clearly visible, but the low frequency noise can also be studied. Inset in (a): the interferences between the reference laser beam $E_{\mathrm{ref}}$ reflecting on the base of the cantilever and the sensing laser beam $E_{\mathrm{sens}}$ on the free end of the cantilever directly gives a calibrated measurement of the deflection $d$, with very high accuracy \cite{Bellon-2008-Patent, Paolino-2009-instrument,Schonenberger-1989,Bellon-2002-OptCom}.}
\label{Fig:Measurement}
\end{figure}

We use BudgetSensors Atomic Force Microscopy (AFM) cantilevers (BS-Cont) with and without gold coating. They present a nominal rectangular geometry: $\SI{450}{\mu m}$ long, $\SI{50}{\mu m}$ wide and $\SI{2}{\mu m}$ thick, with an optional $\SI{70}{nm}$ gold layer on both sides. The measurement is performed with a home made interferometric deflection sensor \cite{Bellon-2008-Patent, Paolino-2009-instrument}, inspired by the original design of Schonenberger \cite{Schonenberger-1989} with a quadrature phase detection technique \cite{Bellon-2002-OptCom}: the interferences between the reference laser beam reflecting on the base of the cantilever and the sensing laser beam on the free end of the cantilever (see inset of Fig.~\ref{Fig:Measurement}(a)) directly gives a measurement of $d$, with very high accuracy.  A first advantage of the technique is that it offers a calibrated measurement of the deflection, without conversion factor from Volt to meter as in the standard optical lever technique common in AFM. We illustrate in Fig.\ref{Fig:Measurement} the performance of our detection system with the power spectrum density of a rigid mirror (bottom black line on each graph): the light intensities on the photodiodes are tuned exactly as during the measurement on the cantilever, but since the mirror is still the measured spectrum reflects only the detection noise. This background noise is as low as $\SI{8e-28}{m^{2}/Hz}$ for the golden coated cantilever measurement, just $\SI{30}{\%}$ higher than the shot noise limit of our detection for a $\SI{200}{\mu W}$ sensing laser beam. At frequencies smaller than $\SI{100}{Hz}$, the electronic noise starts dominating the spectrum but the noise keeps smaller than $\SI{2e-26}{m^{2}/Hz}$ at $\SI{3}{Hz}$. The noise is a bit higher for the raw silicon cantilever measurement, as its reflectivity is smaller and shot noise thus increases.

We plot in Fig.\ref{Fig:Measurement} the PSD of thermal noise driven deflection at the free end of the cantilevers (with and without gold coating) in air at atmospheric pressure. The thermal excitation operates like a white noise force on the cantilever. Within the frequency bandwidth of our acquisition card ($\SI{0-100}{kHz}$), the two resonances present in each spectrum correspond to the first two modes of oscillation of the cantilever modeled as a mechanical embedded-free beam. In agreement with the Euler-Bernoulli model \cite{Graff-1975} these frequencies satisfy the relation $f_1/f_0=6.26$.  We also have a lot of information around the first resonance as the measurements are always above the background noise of the system, notably at low frequency. If we subtract the background noise spectrums from the measurements, we have an estimation of the thermal noise of the first resonance of the system on the whole $\SI{3}{Hz}$ to $\SI{40}{kHz}$ frequency interval, which we can compare to the theoretical expectation of the models presented in the previous section.

\begin{figure}
\begin{center}
\psfrag{PSD}[Bc][Bc]{PSD $S_{d}(f)/\SI{}{(m^2/Hz)}$}
\psfrag{f}[Bc][Bc]{Frequency $f/\SI{}{Hz}$}
\psfrag{(a)}[Bl][Bl]{(a)}
\psfrag{(b)}[Bl][Bl]{(b)}
\psfrag{meas.}[Bl][Bl]{\footnotesize meas.}
\psfrag{SHO}[Bl][Bl]{\footnotesize $\SHO$}
\psfrag{Sader}[Bl][Bl]{\footnotesize $\Sader$}
\psfrag{SHO*}[Bl][Bl]{\footnotesize $\SHO^{*}$}
\psfrag{Sader*}[Bl][Bl]{\footnotesize $\Sader^{*}$}
\includegraphics{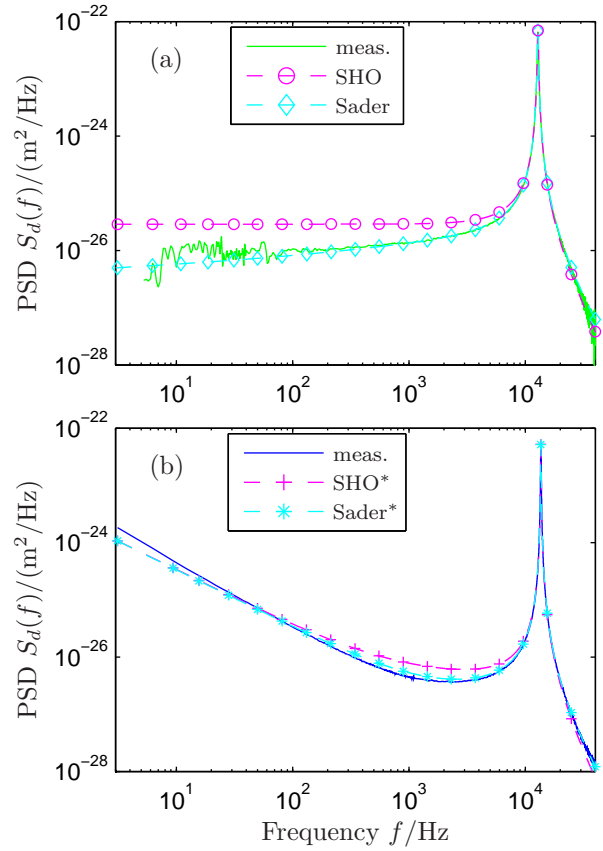}
\end{center}
\caption{Power Spectrum Density (PSD) of thermal noise induced fluctuations for a raw silicon cantilever (a) and a golden coated one (b). Their low frequency behavior are very different: the metallic layer induces a much larger noise when $f$ goes to $0$, with a $1/f$ like frequency dependance that can be modeled with a viscoelastic dissipation (models $\SHO^{*}$ and $\Sader^{*}$). If simple harmonic oscillators models are fine for the resonance, they loose their pertinency at lower frequencies, where the Sader approach stick closer to the experimental observations.}
\label{Fig:Fits}
\end{figure}

The low frequency part of the spectra of the two cantilevers is completely different. For the uncoated cantilever, the trend is a slow decrease as $f$ goes to 0, while the spectrum increases in $1/f$ like fashion when a coating is present. Let us first focus on the raw silicon cantilever. As shown on Fig.\ref{Fig:Fits}(a), if the $\SHO$ model of eq. (\ref{eq:S-SHO}) is fine to fit the resonance, it looses its pertinency at low frequency, where the value of the plateau cannot be tuned without degrading the fit of the resonance. In fact, once matched the resonance frequency ($f_{0}$), width ($\propto 1/Q$) and height ($\propto Q/\kappa$), there are no adjustable parameters left. The $\Sader$ model, introducing a frequency dependance in the viscous dissipation, sticks much closer to the measurement. To compute this prediction, we used the tabulated values for the properties of silicon and air at room temperature, and the explicit formula of ref.\cite{Sader-1998} for the hydrodynamic function. The physical dimensions of the cantilever (length, width and thickness) were tuned within the manufacturer tolerance to match the experimental observation. The agreement is very good till frequencies much smaller than the resonance, and present a validation of Sader's approach outside the resonances where it had been tested up to now \cite{Chon-2000,Maali-2005,Ghatkesar-2008}.

For the coated cantilever, as shown on Fig.\ref{Fig:Fits}(b), viscous damping is clearly inappropriate to interpret the low frequency part of the spectrum. This $1/f$ like behavior was reproducible with cantilevers from various manufacturers and with various metallic coatings, although the stronger effect was obtained with gold. The viscoelastic contribution introduced in the $\SHO^{*}$ and $\Sader^{*}$ model supplies a qualitatively good description of the spectrum. Here a new free parameter is introduced to adjust the data: the loss tangent $\phi$, which is of order $\SI{e-3}{}$. As for the raw cantilever, Sader's approach for viscous dissipation gives better results when we leave the immediate surroundings of the resonance, allowing a better match with the measurement in the few $\SI{}{kHz}$ range. 

To make sure that the low frequency part of the spectrum of the golden coated cantilever doesn't depend on the viscous damping due to the surrounding atmosphere, we made a measurement in vacuum at $\SI{e-5}{mbar}$. Fig. \ref{Fig:or-vide} shows that, in agreement with the viscoelastic models, the only effect of reducing pressure is an increase of the effective quality factor $Q_{\textit{eff}}$, the low frequency part is unchanged and therefore related to an inner behavior of the cantilever. It is also worth mentioning that in vacuum the effective quality factor of the coated cantilever is about one order of magnitude smaller than that of the raw cantilever, pointing also to some higher inner dissipation. This last point has been studied as a function of the gold layer thickness and surrounding atmosphere pressure in ref. \cite{Sandberg-2005}, with similar observations.

A possible artifact in the measurement of such tiny fluctuations is the effect of the light used to sense the deflexion of the cantilever: it has been demonstrated how radiation pressure and local heating by a laser beam can be used to excite a micro-cantilever \cite{Marti-1992}. In the case of interferometric set-up using a resonant cavity between the cantilever surface and an optical fiber, the modulation of light intensity $I$ in the cavity while its length $l-d$ is changing can even induce self oscillation of the mechanical micro-resonator \cite{Kim-2002} or on the contrary damp thermal noise \cite{Metzger-2004}. As the frequency response of the cantilever motion to the input light power is coating dependent \cite{Marti-1992, Sheard-2004}, one may wonder if this could explain the differences at low frequency between raw silicon and golden coated lever. There is anyway a important difference in our setup with respect to the fiber interferometers: there is no resonant cavity in our experiment, and only one reflexion occurs on the cantilever. Unless the intensity is modulated on purpose, there is thus no reason why the cantilever should be oscillated by the sensing light (beside the constant offset due to constant illumination). We checked that our observations at low frequency were not dependent on our measurement system in 3 different ways:
\begin{itemize}
\item We measured the frequency response function $\chi_{dI}(f)$ of the deflexion $d$ to the light intensity $I$ by modulating by a few percent the laser power \cite{Sheard-2004}. In standard measuring conditions, $I$ is kept constant but present small fluctuations around its mean value, which will thus trigger some motion of the cantilever. This light induced deflexion noise $S_{d}^{\mathrm{light}}(f)$ can be estimated using the PSD $S_{I}$ of light intensity and the response function $\chi_{dI}$: $S_{d}^{\mathrm{light}}=|\chi_{dI}|^{2}S_{I}$. It is at least 3 orders of magnitude smaller than the observed thermal noise.
\item On coated cantilevers, if we decrease the interferometer light intensity by a factor of 10, the shot noise will raise, but will still be much lower than the measured thermal noise of the cantilever at low frequency. Except for the raise of the background noise, the PSD $S_{d}(f)$ is not affected by such a drastic change in the laser power, notably at low frequency.
\item We measured the thermal noise of some cantilevers which were coated on one side only. Expect for a lower background noise when measuring on the coated surface (due to a higher reflectivity and thus lower shot noise),  the PSD $S_{d}(f)$ does not change when the cantilever is turned upside down.
\end{itemize}
The $1/f$ like behavior of the thermal noise thus appears to be a robust characteristic of coated cantilevers, which we interpret has the signature of a viscoelastic dissipation process.

\section{Discussion}

\begin{figure}
\begin{center}
\psfrag{Sz}[Bc][Bc]{PSD $S_{d}(f)/\SI{}{(m^2/Hz)}$}
\psfrag{f}[Bc][Bc]{Frequency $f/\SI{}{Hz}$}
\includegraphics{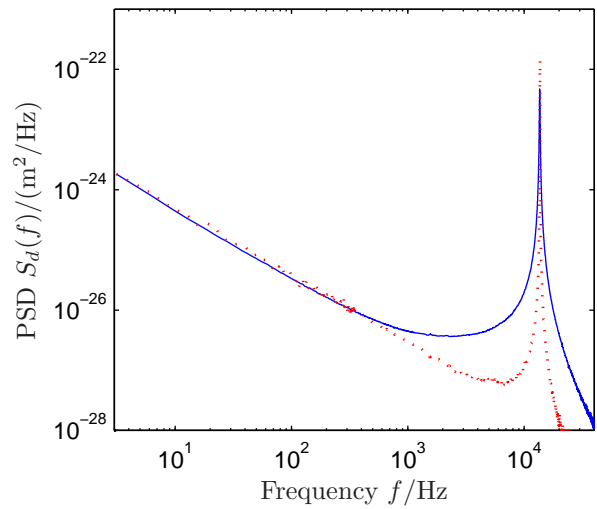}
\end{center}
\caption{Power Spectrum Density (PSD) of thermal noise induced fluctuations for a golden coated cantilever at atmospheric pressure (wide resonance, plain blue) and in vaccum (sharp resonance, dotted red). The low frequency behavior is exactly the same. In agreement with the viscoelastic model, this part of the spectrum doesn't depend on the viscous damping due to the surrounding atmosphere, and is therefore related to a dissipation in the cantilever itself.}
\label{Fig:or-vide}
\end{figure}

In the viscoelastic model we made the implicit hypothesis of a frequency independent complex elastic constant $\kappa^*$. This is an approximation as this can't satisfy the Kramers Kronig relations \cite{deGroot-1984}. In order to improve our understanding of the properties of the system we need to estimate this dependance. We will use for this the Fluctuation-Dissipation Theorem \cite{Callen-1952} and the Kramers-Kronig relations \cite{deGroot-1984}, which allow from  the knowledge of the real or the imaginary part of a transfer function to rebuild entirely this function. Indeed, thanks to the sensitivity of our apparatus, we resolve the power spectrum of fluctuation from very low frequencies to beyond the resonance (Fig. \ref{Fig:or-vide}). Using the FDT (eq. \ref{eq:FDT}), we can estimate from this spectrum the imaginary part of the response function $\Im(1/G(\omega))$. We can consequently rebuilt with an algorithm based on the Kramers Kronig relations the full response function $G(\omega)$ \cite{Schnurr-1997}.

\begin{figure}
\begin{center}
\psfrag{Re(G)}[Bc][Bc]{$\Re(G(\omega)) \ / \ \SI{}{N.m^{-1}}$}
\psfrag{Im(G)}[Bc][Bc]{$\Im(G(\omega)) \ / \ \SI{}{N.m^{-1}}$}
\psfrag{PSD}[Bc][Bc]{PSD $S_{d}(f)\ / \ \SI{}{m^2.Hz^{-1}}$}
\psfrag{f}[Bc][Bc]{Frequency $f/\SI{}{Hz}$}
\psfrag{(a)}[Bc][Bc]{(a)}
\psfrag{(b)}[Bc][Bc]{(b)}
\psfrag{(c)}[Bc][Bc]{(c)}
\psfrag{Exp. (air)}[Bl][Bl]{\small meas. (air)}
\psfrag{Exp (vaccum)}[Bl][Bl]{\small meas. (vac.)}
\psfrag{SHO}[Bl][Bl]{\small $\SHO$}
\psfrag{SHO*}[Bl][Bl]{\small $\SHO^{*}_{\omega}$}
\psfrag{k*}[Bl][Bl]{\small $\kappa^{*}_{\omega}$}
\includegraphics{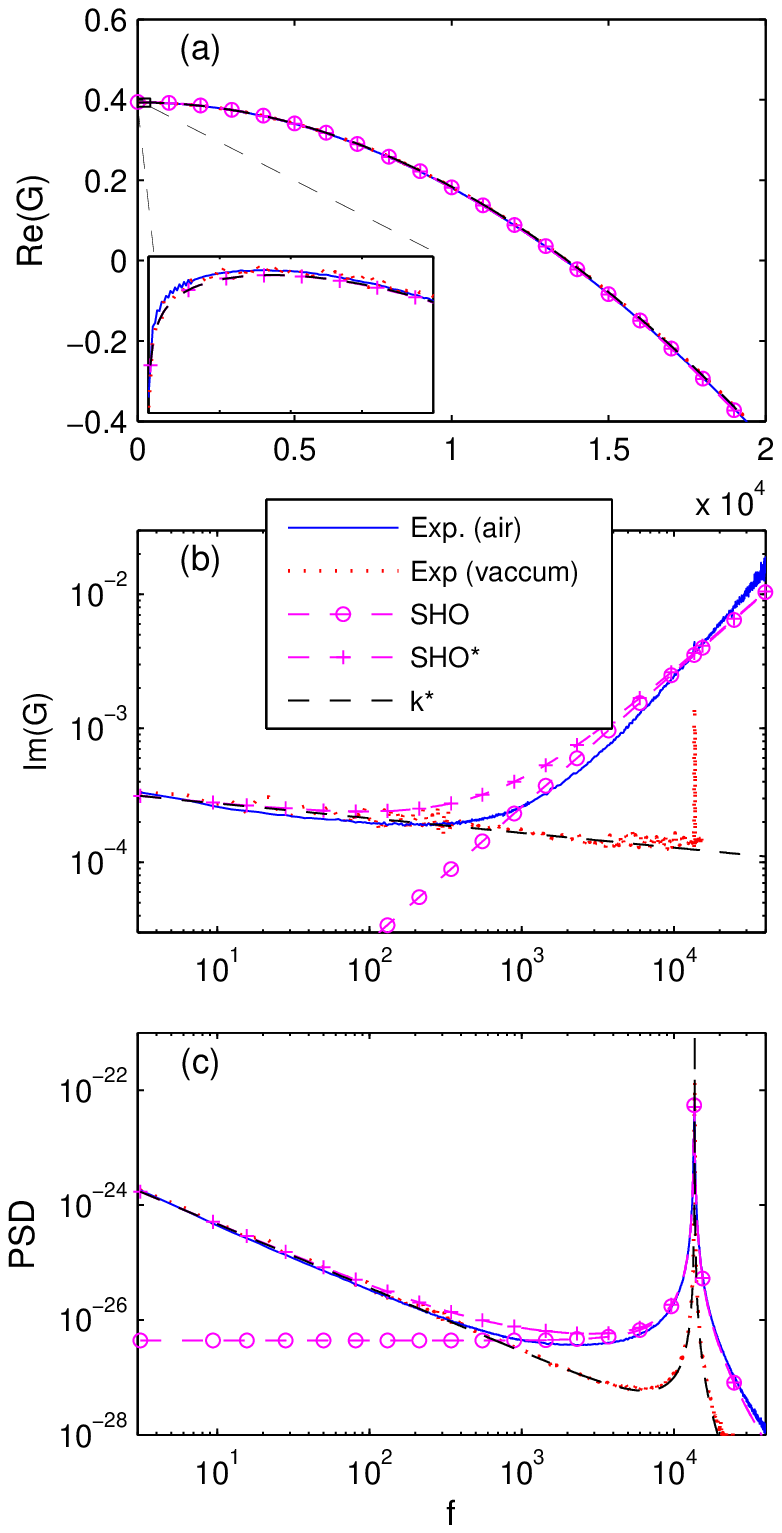}
\end{center}
\caption{Reconstructed mechanical response function and thermal noise spectrum of a golden coated cantilever in air and in vacuum. (a) Real part of the response function.  The inset is a zoom around $f=0$ ($\SI{400}{Hz}$ wide and $\SI{e-3}{N.m^{-1}}$ high). (b) Imaginary part of the response function (log scale on both axes). (c) Power Spectrum Density (PSD) of thermal noise induced fluctuations (log scale on both axes). The viscoelastic dissipation in the cantilever is evidence by the vacuum measurement, and can be fitted by a power law of frequency with a small exponent $\alpha=-0.11$. The low frequency behavior of the real part of $G$ for the viscoelastic models is deduced from Kramers-Kronig relations applied to the power law fit of the imaginary part, it matches well the experimental data. The viscous damping of the $\SHO$ models fails to describe the dissipation off resonance in air.}\label{Fig:exp-SHO}
\end{figure}

We plot in Fig. \ref{Fig:exp-SHO} the result of this reconstruction process for the 2 spectrums of Fig.~\ref{Fig:or-vide}: golden coated cantilever at atmospheric pressure and in vacuum. The $\SHO^{*}$ model provides a qualitative description of the response $G$: the real part is a parabola independent of the value of the pressure, its value at $f=0$ being the spring constant and the quadratic shape resulting from the inertia of the system; while the imaginary part is roughly the addition of a constant term (viscoelasticity) and a linear term in frequency for the measurement in air (viscous damping).

We notice anyway that this model needs some refinement: as expected, the viscoelastic term is not frequency independent. Indeed, in vacuum this is the only dissipative source, and according to eq.~\ref{eq:G-k*} we measure $\Im(G^{\kappa^{*}}(\omega))=\Im(\kappa^{*}(\omega))=\kappa (\omega) \phi (\omega)$. As shown by the fit on vacuum data in Fig.~\ref{Fig:exp-SHO}(b), a very good approximation is given by a power law with a small exponent: $\Im(\kappa^{*}(\omega)) \propto \omega ^{\alpha}$ with $\alpha=-0.11$. This functional form as also the advantage of being compatible with Kramers-Kronig relations, which lead to the full description of the frequency dependent complex spring constant:
\begin{equation}\label{eq:k*_w}
\kappa^{*}(\omega) = \kappa-\kappa_{J} \left( i \frac{\omega}{\omega_{0}}\right) ^{\alpha}
\end{equation}
where $\kappa_{J}$ is real. In dielectric measurements, such frequency dependence has been introduced by Jonscher \cite{Jonscher-1977} and is used to describe a pseudo-conductivity (divergence of the dissipative part of the dielectric constant as $\epsilon'' \propto \omega^{\alpha}$ with $-1<\alpha<0$, pure conductivity corresponding to $\alpha=-1$). The two parameters describing the frequency behavior are extracted from the fit on the imaginary part  $\Im(G^{\kappa^{*}}(\omega))$: for this particular cantilever, we measure $\kappa_{J}=\SI{7e-4}{N.m^{-1}}$ and $\alpha=0.11$.

The zoom at low frequency in the inset of Fig.~\ref{Fig:exp-SHO}(a) shows an unexpected behavior of $\Re(G)$ as $f$ goes to 0: instead of the constant trend predicted by the $\SHO$ and $\Sader$ models, it presents a maximum and a steep decrease (of small amplitude though). At low frequency inertial effects are negligible, thus the real part of $G$ is directly the real part of the spring constant: for frequency dependent viscoelastic models, $\Re(G(\omega))\simeq\Re(\kappa^{*}(\omega))= \kappa-\kappa_{J} \cos(\alpha \pi/2) (\omega/\omega_{0})^{\alpha}$. $\kappa_{J}$ and $\alpha$ are already fixed by the fit on the imaginary part of the reconstructed function $\Im(G(\omega))$, hence we didn't need to introduce any new adjustable parameters to closely match the observations on the real part as illustrated in the inset of Fig.~\ref{Fig:exp-SHO}(a).

\begin{figure}
\begin{center}
\psfrag{Re(G)}[Bc][Bc]{$\Re(G(\omega)) \ / \ \SI{}{N.m^{-1}}$}
\psfrag{Im(G)}[Bc][Bc]{$\Im(G(\omega)) \ / \ \SI{}{N.m^{-1}}$}
\psfrag{PSD}[Bc][Bc]{PSD $S_{d}(f)\ / \ \SI{}{m^2.Hz^{-1}}$}
\psfrag{f}[Bc][Bc]{Frequency $f/\SI{}{Hz}$}
\psfrag{(a)}[Bc][Bc]{(a)}
\psfrag{(b)}[Bc][Bc]{(b)}
\psfrag{(c)}[Bc][Bc]{(c)}
\psfrag{Exp. (air)}[Bl][Bl]{\small meas. (air)}
\psfrag{Exp (vaccum)}[Bl][Bl]{\small meas. (vac.)}
\psfrag{Sader}[Bl][Bl]{\small $\Sader$}
\psfrag{Sader*}[Bl][Bl]{\small $\Sader^{*}_{\omega}$}
\psfrag{k*}[Bl][Bl]{\small $\kappa^{*}_{\omega}$}
\includegraphics{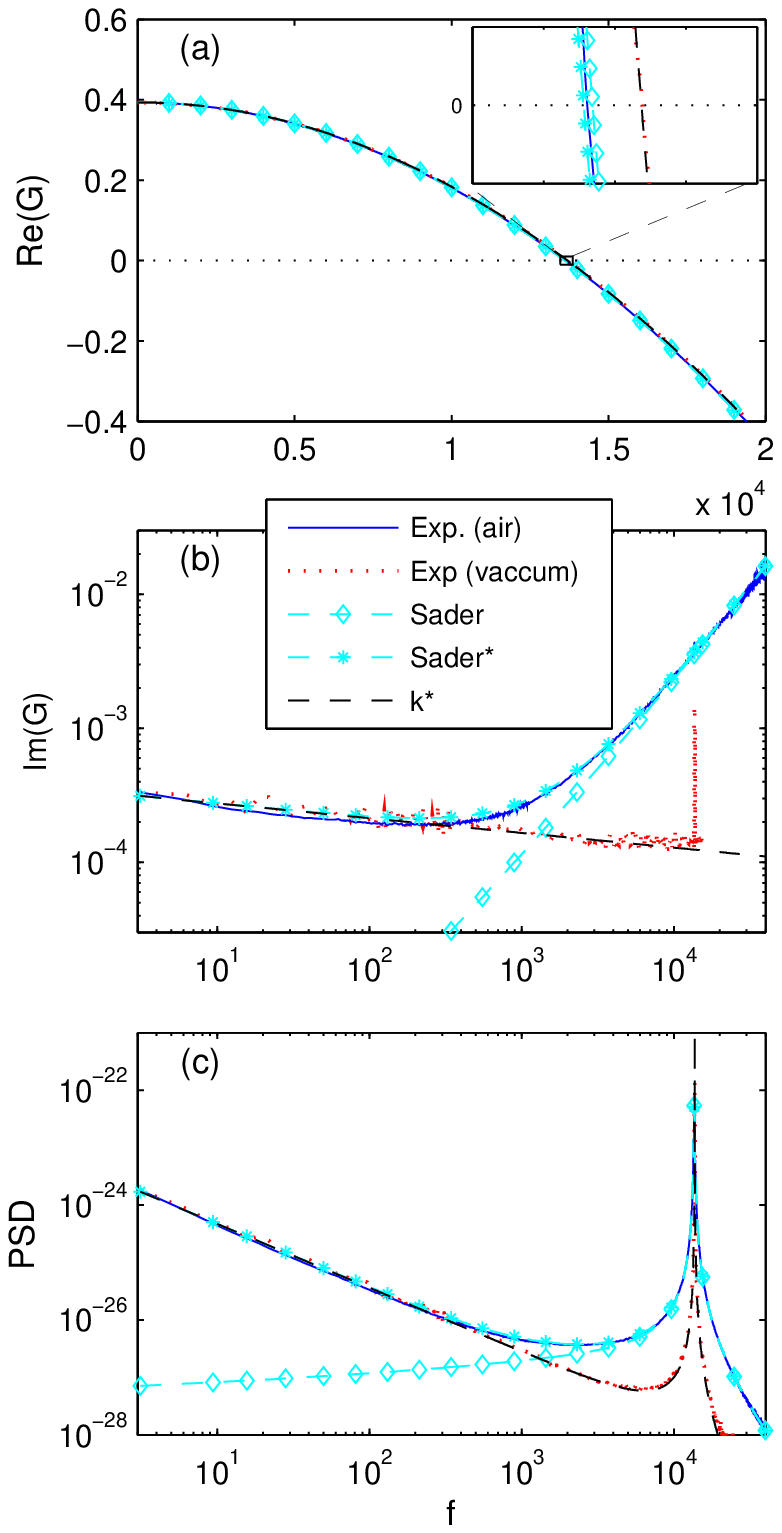}
\end{center}
\caption{Reconstructed mechanical response function and thermal noise spectrum of a golden coated cantilever in air and in vacuum. (a) Real part of the response function. The inset is a zoom around $f=f_{0}$ ($\SI{400}{Hz}$ wide and $\SI{e-3}{N.m^{-1}}$ high). (b) Imaginary part of the response function (log scale on both axes). (c) Power Spectrum Density (PSD) of thermal noise induced fluctuations (log scale on both axes). The Sader models describe accurately the dissipation in air, catching for instance the small frequency shift of the resonance due to the added mass moving along with the cantilever. The $\Sader^{*}_{\omega}$ model, summing the two models for dissipation (frequency dependent viscoelasticity and Sader's), fits adequately the experimental observations in the whole frequency window.}
\label{Fig:exp-Sader}
\end{figure}

From this improved description of the complex spring constant, the models introduced previously can be refined by taking into account the frequency dependence of $\kappa^{*}(\omega)$: we will label as $\kappa^{*}_{\omega}$ the pure viscoelastic model, $\SHO^{*}_{\omega}$ the one adding the simple viscous damping and $\Sader^{*}_{\omega}$ the most complete model including viscoelasticity and Sader's treatment for the surrounding atmosphere. As shown on Fig.~\ref{Fig:exp-SHO}, the $\kappa^{*}_{\omega}$ model gives a very good description of the vacuum measurement and of the reconstructed response function $G$. For the measurement at atmospheric pressure, the $\SHO^{*}_{\omega}$ model is fine at low frequency, but not very accurate to describe data above $\SI{1}{kHz}$: the frequency dependance of the dissipation is not simply linear. This mismatch disappear with the $\Sader^{*}_{\omega}$ model, as illustrated by Fig.~\ref{Fig:exp-Sader}. The agreement is excellent: the frequency dependent viscoelastic dissipation at low frequency and the Sader approach to the viscous coupling with air at high frequency fit very accurately the reconstructed imaginary part of the response function, even in the transition region between the two regimes.

The inset in Fig.~\ref{Fig:exp-Sader}(a) shows how details in the real part are also well described by the model. The zoom around the resonance frequency demonstrates the effect of added mass due to the air around the cantilever: the resonance frequency (corresponding to the zero of $\Re(G)$) is slightly shifted to lower frequency in air with respect to vacuum  ($\SI{-0.5}{\%}$ in frequency corresponding to a $\SI{1}{\%}$ added mass). The Sader models accurately accounts for this observation. 

\section{Conclusions}

We have measured the power spectrum density (PSD) of thermal noise induced deflexion of coated and uncoated micro-cantilevers at atmospheric pressure and down to $\SI{e-5}{mbar}$. Thanks to the sensitivity of our apparatus, we resolve the spectrum completely from very low frequencies to beyond the resonance. The common simple harmonic oscillator model with viscous damping is clearly inadequate to describe the off resonance fluctuations, especially of coated cantilevers. At low frequency, the thermal noise of those shows a $1/f$ like trend, which can be seen as the signature of a viscoelastic dissipation in the cantilever. To go further than simple observations, we use the Fluctuation-Dissipation Theorem and the Kramers-Kronig relations to rebuild from the measured PSD the complete response function of the cantilever. A simple power law is found to describe accurately the frequency dependence of the viscoelastic dissipation, and a consistent model can be proposed to fit tightly all the experimental data: beyond the simple harmonic oscillator approximation, it includes Sader's approach to describe the coupling with the surrounding atmosphere and a mechanical Jonscher like term to account for viscoelasticity.

Let us emphasize a important point here: the use of FDT and Kramers-Kronig relations to rebuilt the mechanical response function of the system is based on very general hypotheses (linear response, causality, thermal equilibrium). Our measurements are thus free of any hypothesis on the dissipation processes in the cantilever, and the viscoelastic model we eventually propose is purely phenomenological. Although viscoelasticity or anelasticity had already been used in several models to account for observations, it has mostly been limited to the resonances of oscillators and to their quality factors. Our experiment offers a complete determination of its properties on a coated micro-cantilever, with quantitative measurement of its amplitude and frequency dependence on a wide spectral range.

The use of thermal noise is a key point of our approach, since we don't need to determine exactly the transfer function of the external forcing method which is usually necessary to measure a response function. We get an excellent resolution with measurement of mechanical loss tangents smaller than $\SI{e-3}{}$. Even such a small dissipation has some consequences on the operation of micro-cantilevers, notably when they are used in vacuum: it gives an upper bound to the quality factor of the resonances. If we don't explain the physical origin of the viscoelasticity due to the coating, we can anyway quantify it and our measurements should be useful in the perspective of testing models of internal friction, eventually leading to improved coating procedures and better performance of cantilever based sensors. Our method would also be suited to study other type of coatings, such as those implied in chemical or biological sensors, alone or linked to the target molecules.

\bigskip

{\bf Acknowledgements}

We thank F. Vittoz and F. Ropars for technical support, and N. Garnier, S. Joubaud, S. Ciliberto, A. Petrosyan and J.P. Aim\'e for stimulating discussions. This work has been partially supported by contract ANR-05-BLAN-0105-01 of the Agence Nationale de la Recherche in France.

\newpage

\bibliographystyle{unsrt}
\bibliography{viscoelastic}

\newpage

\newpage

\newpage

\null

\newpage

\newpage

\newpage

\end{document}